%% file: arxiv.tex
\documentclass[a4paper,USenglish]{article}

\usepackage{amsmath,amssymb,oldlfont,stmaryrd}
\usepackage{colordvi}

\input{all}

\newcommand{\bexample}{\begin{ex}}
\newcommand{\eexample}{\end{ex}}
\newcommand{\twedge}{{\hbox{{\tiny $\wedge$}}}}

\newcommand{\tvee}{{\hbox{{\tiny $\vee$}}}}
\begin{document}

\medskip

\begin{center}
{\Large\bf Implicit complexity via structure transformation}\\[2mm]
Daniel Leivant\fn{Indiana University and IRIF, 
	Universit\'{e} Paris-Diderot}\fn{Research supported by LORIA Nancy
	and by Universit\'{e} de Lyon grant ANR-10-LABX-0070} 
and 
Jean-Yves Marion\fn{LoRIA, Universit\'{e} de Lorraine and CNRS}
\end{center}

\renewcommand{\thefootnote}{\arabic{footnote}}
\setcounter{footnote}{0}

\bigskip

\begin{abstract}
Implicit computational complexity, which aims at characterizing
complexity classes by machine-independent means,
has traditionally been based, on the one hand,
on programs and deductive formalisms for free algebras, and on the other hand
on descriptive tools for finite structures.

We consider here ``uninterpreted" programs for the transformation of
finite structures, which define functions over a free
algebra \dA\ once the elements of \dA\ are themselves considered as finite
structures.
We thus bridge the gap between the two approaches above
to implicit complexity, with the potential of streamlining and clarifying
important tools and techniques, such as set-existence and ramification. 

We illustrate this potential by delineating a broad class of
programs, based on the notion of loop variant familiar
from imperative program construction, 
that characterizes a generic notion of primitive-recursive
complexity, without reference to any data-driven recurrence.
\end{abstract}

\section{Introduction}

{\em Implicit computational complexity (ICC)} 
strives to characterize complexity classes by resource-independent 
methods, thereby elucidating the nature of those classes and relating 
them to more abstract complexity measures, such
as levels of descriptive or deductive abstractions.  
The various approaches to ICC fall, by and large, into two broad classes.
One is {\em descriptive complexity,}
which focuses on finite structures, and as such
forms a branch of Finite Model Theory  \cite{Immerman-Desc}.
Its historical roots go back at least to
the characterization of Log-Space queries by recurrence \cite{Hartmanis72},
and of NP by existential set-quantification \cite{Fagin74}.

The other broad class in ICC focuses on
computing over infinite structures,
such as the natural numbers, strings, or lists,
and uses programming and proof-theoretic methods 
to articulate resource-independent characterizations of complexity classes.
 
We argue here that computing over finite structures
is, in fact, appropriate for implicit complexity 
over {\em infinite} structures as well. 
Our point of departure is the observation that inductive data-objects,
such as natural numbers, strings and lists, are themselves finite 
structures, and that
their computational behavior is determined
by their internal makeup
rather than by their membership in this or that infinite structure.
For example, the natural number three is the structure (or more precisely
{\em partial-}structure, see below) 
$$
\calT(3) \quad \equiv \quad\quad 
\ttzero \,
        \circ \ara{\ttscript{s}}
        \circ \ara{\ttscript{s}}
        \circ \ara{\ttscript{s}}
        \circ \zero
$$
Lifting this representation, a function $f:\, \dN \sra \dN$ 
is perceived as a mapping over finite second-order objects,
namely the natural numbers construed as structures.
This view of inductive objects as finite structures
is implicit already in long-standing representations, such as
the Church-Berarducci-B{\"{o}}hm lambda-coding of inductive data
\cite{Church41,BohmB85}.

As a programming language of reference we propose a Turing-complete 
imperative language \bxbf{ST} for structure transformation,
in the spirit of Gurevich's ASMs \cite{Borger02,Gurevich93e,Gurevich01}.
We regard such programs as operating over classes of finite structures.


We illustrate the naturalness and effectiveness of our approach
by delineating a variant \bxbf{STV} of \bxbf{ST}, based on the notion of 
{\em loop variants} familiar from program development and verification
\cite{Gries81,Dijkstra76,Winskel93}, and proving that it captures exactly
primitive recursion, in the strongest possible sense:
all functions defined by recurrence over free algebras are computable
directly by \bxbf{STV} programs, 
and all \bxbf{STV} programs run in time and space
that are primitive-recursive in the size of the input.

We caution against confounding our approach with
unrelated prior research addressing somewhat similar themes.
Recurrence and recursion over finite structures have been shown 
to characterize logarithmic space and polynomial time queries,
respectively \cite{Hartmanis72,Sazonov80}, but the programs in
question do not allow inception of new structure elements, and so
remain confined to linear space complexity, and are inadequate for 
the kind of characterizations we seek.
On the other hand, unbounded recurrence over arbitrary structures 
has been considered by a number of authors 
\cite{AndaryPV05,AndaryPV11,StrahmZ08},
but always in the traditional sense of computing within an infinite structure.
Also, 
while the meta-finite structures of \cite{GradelG94} merge finite and
infinite components, both of those are
considered in the traditional framework, whereas we deal with purely finite
structures, and the infinite appears via the
consideration of {\em collections} of such structures. 
Finally, the functions we consider are 
from structures to structures (as in \cite{Sazonov80}), 
and are thus unrelated to the global functions of
\cite{Gurevich88, Ebbinghaus-Flum-FMT}, which are
(isomorphism-invariant) mappings that assigns to each structure
a function over it. 


\section{General setting}

\subsection{Partial structures}

We use the phrase {\em vocabulary} for a finite set $V$ of
function-identifiers and relation-identifiers,
with each identifier \ttI\ assigned an arity $\geq 0$ denoted $\ar(\ttI)$.
We refer to nullary function-identifiers as {\em tokens},
and to ones of arity 1 as {\em pointers.}

By {\em $V$-structure} we'll mean here a finite partial-structure
over the vocabulary $V$; that is, a $V$-structure \calS\ consists of
a finite non-empty universe $|\calS|$,
for each function identifier \bff\ of $V$ 
a {\em partial}-function 
$\bff_{{\tinycalS}}: \, |\calS|^k \pa |\calS|$, where $k = \ar(\bff)$,
and for each relation-identifier \bfQ\ of $V$, a relation
$\bfQ_{{\tinycalS}} \subseteq |\calS|^k$, where $k= \ar(\bfQ)$.
We refer to the elements of $|\calS|$ as \calS's {\em nodes.}

We insist on referring to partial-structures
since we consider partiality to be a core component of our approach.
For example, we shall
identify each string in $\{\ttzero,\ttone\}^*$ with a structure over the
vocabulary with a token $\tte$ and pointers
\ttzero\ and \ttone.  So \bxtt{011}
is identified with the four element structure
$$
\tte \,
	\circ \ara{\ttscript{0}}
	\circ \ara{\ttscript{1}}
	\circ \ara{\ttscript{1}}
	\circ \zero
$$
Here \ttzero\ is interpreted as the partial-function defined only
for the leftmost element, and \ttone\ as the partial-function
defined only for the second and third elements.

We might, in fact, limit attention to
vocabularies without relation identifiers,
since a $k$-ary relation $Q$ ($k > 0$) can be represented by its {\em support,}
that is the $k$-ary partial-function
$$
\bgrs(x_1, \ldots , x_k) \df \bxbf{ if } \vec{x} \in Q \bxbf{ then } x_1
        \bxbf{ else } \text{undefined}
$$
Thus, for instance, $Q$ is empty iff $\bgrs$ is empty (which is not the
case if relations are represented by their {\em characteristic} functions).
Note that by using the support rather than the characteristic function
we bypass the traditional representation of
truth values by elements, and obtain
a uniform treatment of functional and relational
structure revisions (defined below), as well as initiality conditions.

A tuple of structures is easily presentable
as a single structure.
Given structures $ \calS_1 , \, \ldots \, , \calS_k$, where $\calS_i$ is
a $V_i$-structure,
let $V$ be the disjoint union of $V_1, \ldots, V_k$,
and let $\calS_1 \oplus \calS_2 \oplus \cdots \oplus \calS_k$
be the $V$-structure whose universe is the disjoint
union of $|\calS_i|$ ($i=1..k$), and where the interpretation
of an identifier of $V_i$
is the same as it is in $\calS_i$, i.e.\ is empty/undefined on $|\calS_j|$ for
every $j \neq i$.

\subsection{Accessible structures and free structures}
The {\em terms over $V$,} or {\em $V$-terms,}
are generated by the closure condition: if $\ar(\bff) = k$
and $\gra_1 \ldots \gra_k$ are terms, then so is 
$\bff\,\gra_1 \cdots \gra_k$. (We use parentheses and commas for
function application only at the discourse level.)
Note that we do not use variables, so our ``terms'' are all closed. 
The {\em height} $\hgt(\bft)$ of a term \bft\ is the height of
its syntax-tree: $\hgt(\bff \gra_1 \cdots \gra_k) =
	1 + \max\{\hgt(\gra_i)| i=1..k\}$.
Given a $V$-structure \calS\ the {\em value} of a $V$-term \gra\ in \calS,
$\lsem \gra \rsem_{{\tinycalS}}$, is defined as usual by recurrence on \gra:
If $\ar(\bff) = k$, then $\lsem \bff \gra_1 \cdots \gra_k \rsem_{{\tinycalS}}
	= \bff_{{\tinycalS}}(\lsem \gra_1 \rsem_{{\tinycalS}},
		\ldots , \lsem \gra_k \rsem_{{\tinycalS}})$
We say that a term \gra\ {\em denotes} its value $v$, and also that it is an {\em address}
for $v$.

A node of a $V$-structure \calS\ is {\em accessible}
if it is the value in \calS\ of a $V$-term.
The {\em height} of an accessible node
$a$ is the minimum of the heights of addresses of $a$.
A structure \calS\ is {\em accessible} when all its nodes are accessible.
If, moreover, every node has
a unique address we say that \calS\ is {\em free}.

A $V$-structure \calT\ is a {\em term-structure} if
\be
\li its universe consists of $V$-terms; and 
\li if $\bff \gra_1,\ldots, \gra_k \; \in |\calT|$ then 
$\gra_1,\ldots, \gra_k \, \in |\calT|$ and
$\bff_{{\tinycalS}}(\gra_1 , \ldots ,\gra_k) = \bff\gra_1 \cdots \gra_k$.
\ee
From the definitions we have
\bprop
A $V$-structure \calS\ is free iff it is isomorphic to a term $V$-structure.
\eprop

Note that if $V$ is functional (no relation identifiers), then 
for each $V$-term \bfq\ we have a free term-structure 
$\calT(\fnbfq)$ consisting of the sub-terms of \bfq\ (\bfq\ included).
Each $\calT(\fnbfq)$ can be represented as a dag of terms,
whose terminal nodes are tokens. 
It will be convenient to fix a reserved token,
say $\bullet$, that will denote in each structure $\calT(\fnbfq)$ 
the term \bfq\ as a whole.

\section{Structure-transformation programs}

Programs operating on structures and transforming them
are well known, for example from Gurevich's Abstract State Machines
\cite{Gurevich93e,Gurevich01,Borger02}.  We define a version of such programs,
giving special attention to basic execution steps (structure revisions).

\subsection{Structure revisions}
We consider the following basic operations on $V$-structures,
transforming a $V$-structure \calS\ to a $V$-structure $\calQ$
which, aside from the changes indicated below,
is identical to \calS.

\bi
\li\ul{\bxsl{Function-revisions}} \threemm
\be
\li A {\em function-extension} is an expression
$\bff\, \gra_1 \cdots \gra_k \downarrow \grb$.
The intent is that if\\
 $\lsem\gra_1\rsem_{{\tinycalS}}, \ldots,
	\lsem\gra_k\rsem_{{\tinycalS}},\lsem\grb\rsem_{{\tinycalS}}$ \quad
are all defined, but \quad $\lsem\bff(\gra_1, \ldots, \gra_k)\rsem_{{\tinycalS}}$
\quad is undefined,\\ then \quad
$\bff_{{\tinycalQ}}(\lsem\gra_1\rsem_{{\tinycalS}}, \ldots,
        \lsem\gra_k\rsem_{{\tinycalS}}) = \lsem\grb\rsem_{{\tinycalS}}$.
\bff\ is the {\em eigen-function} of the extension.

\li A {\em function-contraction} is an expression
$\bff\gra_1 \cdots \gra_k \suparrow$.
The intent is that\\ 
$\bff_{{\tinycalQ}}(\lsem\gra_1\rsem_{{\tinycalS}}, \ldots,
        \lsem\gra_k\rsem_{{\tinycalS}})$ is undefined.
\ee

\li \ul{\bxsl{Relation-revisions}}

Relation revisions may be viewed as a special case of function-revisions,
given the functional representation of relations described above.
We mention them explicitly since they are used routinely.
\be
\li A {\em relation-extension} is an expression
$\bfR \sdownarrow (\gra_1, \ldots, \gra_k)$
where \bfR\ is a $k$-ary relation identifier.
The intent is that if each $\lsem\gra_i\rsem_{\tinycalS}$ is defined,
then $\lsem \bfR\rsem_{\tinycalQ}$ is $\lsem\bfR\rsem_{\tinycalS}$
augmented with the tuple $\lng \lsem\gra_1\rsem_{\tinycalS}, \ldots, 
 	\lsem\gra_k\rsem_{\tinycalS}\rng$ (if not already there).
\bfR\ is the {\em eigen-relation} of the extension.
\li A {\em relation-contraction} is an expression
$\bfR \suparrow (\gra_1, \ldots, \gra_k)$.
The intent is that if each $\lsem\gra_i\rsem_{\tinycalS}$ is defined,
then $\lsem \bfR\rsem_{\tinycalQ}$ is $\lsem\bfR\rsem_{\tinycalS}$
with the tuple $\lng \lsem\gra_1\rsem_{\tinycalS}, \ldots, 
 	\lsem\gra_k\rsem_{\tinycalS}\rng$ removed (if there).
\ee

\li \ul{\bxsl{Node-revisions}}
\be
\li A {\em node-inception} is an expression of the form 
$\bfc \ssDownarrow$, where \bfc\ is a token.
The intent is that, if $\lsem\bfc\rsem_{{\tinycalS}}$ is undefined,
then $|\calQ|$ is $|\calS|$ augmented with a new node \grn\ denoted by \bfc\
(i.e.\ $\lsem\bfc\rsem_{{\tinycalQ}} = \grn$).
A traditional alternative notation
is $\bfc := \bxbf{new}$. Assigning \grn\ to a compound
address $\bff\gra_1\cdots\gra_k$ can be viewed as an abbreviation for
$\bfc \sDownarrow; \;\;\bff\gra_1\cdots\gra_k \sdownarrow \bfc; \;\;
\ttc\suparrow$, where \bfc\ is a fresh token.

\li  A {\em node-deletion} is an expression of the form 
$\bfc\sUparrow$, where \bfc\ is a token.
The intent is that  
$\calQ$ is obtained from \calS\ by
removing the node $\grn = \lsem\bfc\rsem$ (if defined),
and removing all tuples containing \grn\ from each $\bfR_{{\tinycalS}}$
(\bfR\ a relation-identifier)  and from the graph of
each $\bff_{\tinycalS}$ (\bff\ a function identifier).
Again, a more general form of node-deletion,
$\bff\vec{\gra}\sUparrow$, can be implemented as the 
composition of a function-extension $\bfc \sdownarrow \bff\vec{\gra}$ 
and $\bfc\sUparrow$, \bfc\ a fresh token.

Deletions are needed, for example, when the desired output structure
has fewer nodes than the input structure (``garbage collection'').
\ee
\ei
We refer to the operations above collectively as 
{\em revisions}. 
Revisions cannot be split into smaller actions.
On the other hand, a function-extension and a function-contraction
can be combined into an
{\em assignment}, i.e.\ a phrase of the form $\bff\vec{\gra} := \grb$.
This can be viewed as an abbreviation, with \ttb\ a fresh token, 
for the composition of four revisions:
$$
\ttb \sdownarrow \grb; \;\;
\bff\vec{\gra}\uparrow; \;\;
\bff\vec{\gra} \sdownarrow \ttb; \;\;
\ttb\suparrow
$$



\subsection{\bxbf{ST} Programs}

Our programming language \bxbf{ST} consists of guarded iterative programs
built from structure revisions.
Uninterpreted programs over a vocabulary $V$ normally refer to an expansion 
$W$ of $V$, as needed to implement algorithms and to generate output.
We refer from now to such an expansion $W$.

\bi
\li 
A {\em test} is one of for following types of phrases.
\be
\li A {\em convergence-expression} $! \gra$, where \gra\ is an address.
This is intended to state that the address \gra\ is defined for the current
values of the function-identifiers. Thus $\neg ! \gra$
states that \gra\ is undefined in the current structure.

\li An {\em equation} $\gra \sequal \grb$ where \gra\ and \grb\
are addresses.  This is intended to state that both addresses are defined and
evaluate to the same node.

\li A {\em relational-expression} $\bfR\gra_1\cdots\gra_k$,
where $\bfR$ is a $k$-ary relation-identifier and each $\gra_i$ is an address.
By the convention above, this may be construed as a 
special case of the equation $\grs_{\fnbfR}\gra_1\cdots\gra_k = \gra_1$.
\ee

\li A {\em guard} is a boolean combination of tests.
\ei

Given a vocabulary $V$ the {\em $V$-programs of \bxbf{ST}}
are generated inductively as follows (we omit the reference
to $V$ when un-needed). 
\be
\li A structure-revision is a program.
\li If $P$ and $Q$ are programs then so is $P;\,Q$.
\li If $G$ is a guard and $P,Q$ are programs, 
then $\;\bxbf{if}\,[G]\,\{P\}\,\{Q\}\;$ and
$\; \bxbf{do}\;[G]\,\{P\}\;$ are programs.
\ee

\subsection{Program semantics}

Given a vocabulary $V$,
a {\em $V$-configuration (cfg)} is a $V$-structure.
Given a $V$-structure \calQ\ and $W \supseteq V$,
we write $\calQ^{W}$ for the expansion of \calQ\ to $W$ with
all identifiers in $W \sminus V$ interpreted as empty
(everywhere undefined functions and empty relations).
For a program $P$ over $V$ we define
the binary {\em yield relation} $\rA_P$ between 
$V$-configurations by recurrence on $P$.
For $P$ a structure-revision the definition follows the intended semantics
described informally above.  The cases for composition, branching,
and iteration, are straightforward as usual.

Let $\grF:\; \gothC \pa \gothC'$ be a partial-mapping from a class
\gothC\ of $V$-structures to a class $\gothC'$ of $V'$-structures.
A $W$-program $P$ {\em computes \grF} if for every
$\calS \in \gothC$, $\calS \rA_P \calQ$ for some $W$-expansion \calQ\
of $\grF(\calS)$.

The vocabulary $V'$ of the output structure
$\calT$ need not be related to the input vocabulary $V$.\fn{Of 
course, if \gothC\ is a proper class 
(in the sense of G\"{o}del-Bernays set theory), then
the mapping defined by $P$ is a proper-class.} 

We shall focus mostly on programs as transducers.
Note that all structure revisions refer only to accessible structure nodes.
It follows that non-accessible nodes play no role in the computational
behavior of \bxbf{ST} programs.  We shall therefore focus
from now on accessible structure only.

\subsection{Examples}\label{subsec:examples}

\be
\li \bxbf{Concatenation by splicing.} 
The following program computes concatenation over $\{\ttzero,\ttone\}^*$.
It takes as input a pair $\calT)u) \oplus \calT(v)$ of structures, 
where the nil and two successor identifiers are 
$\tte,\ttzero,\ttone$ for $\calT(u)$ and $\hat{\tte},\hat{\ttzero},\hat{\ttone}$
for $\calT(v)$.
The output is $\calT(u \cdot v)$, with vocabulary $\tte,\ttzero,\ttone$.

$$\begin{array}{lll}
\tta \downarrow \tte; & \bxsl{\% moving token \tta\ to end of input 1} \\[1mm]
\bxbf{do}\;[!\, \ttzero\tta \; \vee \; !\, \ttone\tta]\;\\[1mm]
\zero\qquad \{\; \tta  :=  \ttzero\tta ; 
	\; \tta  := \ttone\tta \}; 
		& \bxsl{\% note: only one of \ttzero\tta, \ttone\tta\ is defined}
\\[1mm]
\ttzero\tta \downarrow \hat{\ttzero}\hat{\tte} ; \;
	\ttone\tta \downarrow \hat{\ttone}\hat{\tte};
	\tta := \ttzero\tta; \; \tta := \ttone\tta;
		& \bxsl{\% hooking $\ttzero/\ttone$ to input 2}\\[1mm]
\bxbf{do}\;[!\, \hat{\ttzero}\tta \; \vee \; !\, \hat{\ttone}\tta] &
	\bxsl{\% copying input 2 to $\ttzero/\ttone$} \\[1mm]
\zero\qquad
 		\{ 
		\; \ttzero\tta \downarrow \hat{\ttzero}\tta;
 		\; \ttone\tta \downarrow \hat{\ttone}\tta;
		\; \tta \downarrow  \ttzero\tta;
		\; \tta \downarrow  \ttone\tta  \; \}
\zero\qquad
\end{array}$$

\li \bxbf{Concatenation by copying.}
The previous program uses no inception, as it splices the second argument over the first.
The following program copies the second argument over the first,
thereby enabling a repeated and modular use of concatenation, as in
the multiplication example below.

$$\begin{array}{ll}
\tta \downarrow \tte; &
	\bxsl{\% moving \tta\ to end of input 1} \\[1mm]
\bxbf{do}\;[\, !\, \ttzero\tta \; \vee \; !\, \ttone\tta\, ]\;\\
\zero\qquad \{\; \tta  := \ttzero\tta ;\;  
	\tta := \ttone\tta \; \};\\[1mm]
\ttb \downarrow \hat{\tte};
	& \bxsl{\% copy of input 2 incepted after input 1}\\[1mm]
\bxbf{do}\;[!\, \hat{\ttzero}\ttb \; \vee \; !\, \hat{\ttone}\ttb]\\[1mm]
\zero\qquad \{ \ttc \Downarrow;\\[1mm]
\zero\qquad \bxbf{if} \; [\, !\hat{\ttzero}\ttb\, ]\\[1mm]
\zero\qquad \zero\qquad \{\; \ttzero\tta \downarrow \ttc; 
		\; \tta \downarrow \ttc;
		\; \ttb := \hat{\ttzero}\ttb \;\}\\[1mm]
\zero\qquad \zero\qquad \{\; \ttone\tta \downarrow \ttc;
		\; \tta \downarrow \ttc;
		\; \ttb := \hat{\ttone}\ttb \; \};\\[1mm]
\zero\qquad \ttc \suparrow\\[1mm]
\zero\qquad \}
\end{array}$$


\li \bxbf{String multiplication} is the function that
for inputs $w \in \{\ttzero,\ttone\}^*$ and $n \in \dN = \{\tts\}^*$ 
returns $nw =$ the result of concatenating $n$ copies of $w$. 
This is computed by the following program, which takes
as input a pair $\calT(n) \oplus \calT(w)$ of structures, 
with vocabularies $\{\ttz,\tts\}$ and $\{\tte,\ttzero,\ttone\}$ respectively,
and output vocabulary $\{\tte,\ttzero,\ttone\}$.

$$\begin{array}{ll}
\tti \downarrow \ttz;\; \tta \downarrow \tte; \;\\[1mm]
\bxbf{do}\;[\; !\, \tts\tti\; ]
	& \bxsl{\% iterate over numerical input}\\[1mm]
\zero\qquad \{\; \tti  :=  \tts\tti;\\[1mm]
\zero\qquad \; \ttb := \tte;
	& \bxsl{\% copy of string input concatenated over} \\[1mm]
\zero\qquad \bxbf{do}\;[\; ! \ttzero\ttb \; \vee \; ! \ttone\ttb\; ] \\[1mm]
\zero\qquad\qquad \{ \ttc \sDownarrow;\\[1mm]
\zero\qquad\qquad  \bxbf{if} \; [!\ttzero\ttb]\\[1mm]
\zero\qquad\qquad\qquad \{ \ttzero\tta \downarrow \ttc ;\; \ttb := \ttzero\ttb \}\\[1mm]
\zero\qquad\qquad\qquad \{ \ttone\tta \downarrow \ttc ;\; \ttb := \hat{\ttone}\ttb\};\\[1mm]
\zero\qquad\qquad \tta := \ttc ;\; \ttc \suparrow\\[1mm]
\zero\qquad\qquad \} \\[1mm]
\zero\qquad \}
\end{array}$$

\ee

\subsection{Computability}

Since guarded iterative programs are well known to be sound and complete 
for Turing computability, the issue of interest here is
articulating Turing computability in the \bxbf{ST} setting.
Consider a Turing transducer over an I/O alphabet \grS,
with full alphabet $\grG \supset \grS$, set of states $Q$, start state
\tts, print state \ttp, and transition function \grd.
The input $w = \grs_1 \cdots \bfs_k \in \grS^*$ is taken to be the
structure 
$\tte \circ \ara{\sigma_1} \circ \cdot\cdot\cdot \circ \ara{\sigma_k} \circ$.
  
Define $V_M$ to be the vocabulary with
$\tte$, \ttc\ and each state in $Q$ as tokens; and with
\ttr\ and each symbol in \grG\ as pointers.
Thus the program vocabulary is broader than the input vocabulary,
both in representing $M$'s machinery, and with auxiliary components.
The intent is that a configuration 
$(q, \grs_1 \cdots \ul{\sigma}_i \, \cdots \grs_k)$ (i.e.\ with
$\ul{\sigma}_i$ cursored)
be represented by the $V_M$-structure
$$ \begin{array}{rccccccccc}
& \circ & \!\!\!\! \ara{\sigma_1} \,\,\, \circ & \cdot\cdot\cdot &
	& \circ & \!\!\!\! \ara{\sigma_i} \,\, \circ & \cdot\cdot\cdot &
		\circ \; \ara{\sigma_k} \; \circ\\[-1mm] \;
& \tte,\ttq & & && \ttc &
\end{array}$$
All remaining tokens are undefined.

The program simulating $M$ implements the following phases:
\be
\li Convert the input structure into the structure for the initial
configuration, and initialize \ttc\ to the initial input element,
and \ttr\ to be the destructor function for the input string.
\li Main loop: configurations are revised as called for by \grd.
The pointer \ttr\ is used to represent backwards cursor movements.
The loop's guard is \ttp\ (the ``print" state) being undefined.
\li Convert the final configuration into the output.
\ee 

\section{\bxbf{STV}: programs with variants}

\subsection{Loop variants}





A {\em variant} 
is a finite set $T$ of function- and relation-identifiers of positive arity,
to which we refer as $T$'s {\em components.}

Given a vocabulary $V$ the {\em $V$-programs of \bxbf{STV}}
are generated inductively as follows, in tandem with the notion
of a variant $T$ being {\em terminating} in an \bxbf{STV}-program $P$.
Again, we omit the reference to $V$ when it is clear or irrelevant.
\be
\li A structure-revision over $V$ is a program.
A variant $T$ is {\em terminating} in any revision except for
a function- or relation-extension whose eigen function/relation is
a component of $T$.

\li If $P$ and $Q$ are \bxbf{STV}-programs with $T$ terminating,
then so is $P;\,Q$.

\li If $G$ is a guard and $P,Q$ are \bxbf{STV}-programs with $T$ terminating,
then so is $\;\bxbf{if}\,[G]\,\{P\}\,\{Q\}\;$.

\li If $G$ is a guard, and $P$ is a \bxbf{STV}-program with $S$ and 
$T$ terminating variants, then $\; \bxbf{do}\;[G]\,[S]\, \{P\}\;$ 
is a \bxbf{STV}-program, with $T$ terminating.
\ee

We write \bxbf{STV}$(W)$ for the programming language 
consisting of \bxbf{STV}-programs over vocabulary $W$, and omitting $W$ when
in no loss of clarity.


\subsection{Semantics of \bxbf{STV}-programs}

The semantics of \bxbf{STV}-programs is defined as for programs of \bxbf{ST},
with the exception of the looping construct \bxbf{do}.
A loop  $\bxbf{do}\;[G]\,[T]\, \{P\}$ is entered 
if $G$ is true in the current state, and is re-entered
if $G$ is true in the current state, {\em and} the previous pass executes at 
least one contraction for some component of the variant $T$.
Thus, as $\bxbf{do} \, [G][T]\,\{P\}$ is executed, no component of $T$
is extended within $P$ (by the syntactic condition that $T$
is terminating in $P$),
and is contracted at least once for each iteration,
save the last (by the semantic condition on loop execution).

\subsection{String duplication}\label{subsec:vexamples}

The following program duplicates
a string given as a structure: the output structure has the same nodes
as the input, but with functions appearing in duplicate.
The algorithm has two phases: a first loop, with the variant consisting
collectively of the functions,
creates two new copies of the string (while depleting the input function 
in the process). A second loop 
restores one of the two copies to the original identifiers, 
thereby allowing the duplication to be useful within a larger program 
that refers to the original identifiers.
Function duplication in arbitrary structures is more complicated, and
will be discussed below.

$$\begin{array}{ll}
\zero\qquad \tta:= \tte;\\[1mm]
\zero\qquad \bxbf{do}\;[!\ttzero\tta
                \;\vee\; !\ttone\tta]\, [\ttzero,\ttone]
		& \text{\% $\ttzero/\ttone$ copied to 
		  $\bar{\ttzero}/\bar{\ttone}$ 
			and $\hat{\ttzero}/\hat{\ttone}$}  \\[1mm]
\zero\qquad\qquad   \{ \; \ttb \downarrow \tta;
		& \quad \text{while being consumed as variant}\\[1mm]
\zero\qquad\qquad        \bxbf{if}\; [\, !\ttzero\tta\, ]\\[1mm]
\zero\qquad\qquad \;\;    \{ \; \bar{\ttzero}(\tta) \downarrow \ttzero\tta;
		\; \hat{\ttzero}(\tta) \downarrow \ttzero\tta;
		\; \tta := \ttzero\tta;
		\; \ttzero\ttb \uparrow \, \}\\[1mm]
\zero\qquad\qquad \;\;  \{ \; \bar{\ttone}(\tta) \downarrow \ttone\tta;
		\; \hat{\ttone}(\tta) \downarrow \ttone\tta;
		\; \tta := \ttone\tta;
		\; \ttone\ttb \uparrow \, \} \\[1mm]
\zero\qquad\qquad \}; \\[1mm]
\zero\qquad \tta:= \tte; 
	& \text{$\hat{\ttzero}/\hat{\ttone}$ restored to $\ttzero/\ttone$}\\[1mm]
\zero\qquad \bxbf{do}\;[\; !\hat{\ttzero}\tta
                \;\vee\; !\hat{\ttone}\tta \;]\; 
		[\,\hat{\ttzero},\hat{\ttone}\,]\\[1mm]
\zero\qquad \qquad \{ \bxbf{if} \; [\; !\hat{\ttzero}\tta\;]\\[1mm]
\zero\qquad\qquad \quad  \{ \; \ttzero\tta \sdownarrow \hat{\ttzero}\tta;
			\; \hat{\ttzero}\tta \suparrow;
			\; \tta := \ttzero\tta;
			\;\}\\[1mm]
\zero\qquad\qquad \quad  \{ \; \ttone\tta \sdownarrow \hat{\ttone}\tta;
			\; \hat{\ttone}\tta \suparrow;
			\; \tta := \ttone\tta;
			\;\}\\[1mm]
\zero\qquad\qquad \}
\end{array}$$

The ability of \bxbf{STV} programs to duplicate structures (for now only
string structures) is at the core their ability to implement recurrence,
so be discussed below.  

\subsection{Further examples}

\be
\li \bxbf{Concatenation.} 
Using string duplication, we can easily convert the concatenation
examples of \S\ref{subsec:examples} to \bxbf{STV}.
The changes are similar for the splicing and for the
copying programs.  The programs
are preceded by the duplication of each of the two inputs.
The copy of $\ttzero,\ttone$ is then used as guard for the first loop,
and is depleted by an entry in each cycle.
The copy of $\hat{\ttzero},\hat{\ttone}$ is used as guard for the second loop,
and is similarly depleted.

\li \bxbf{Multiplication.}
The program of \S\ref{subsec:examples} is preceded by a duplication of
the string input.  The outer loop has \tts\ as a variant, which is 
depleted by a contraction in each cycle of the current $\tts\tti$.
The inner loop has the copy of $\ttzero,\ttone$ as variant.

\li \bxbf{Exponentiation}
A program transforming the structure for $\ttone^{[n]}\tte$
to the structure for $\ttone^{2^n}$ is obtained by combining the programs for
duplication and concatenation. Using for the input vocabulary
a token \ttz\ and a pointer \tts, and for output a token \tty\ and
a pointer \ttt, The program first initializes the output to
the structure for 1. The main loop has $!\tts\tte$ as guard
and \tts\ as variant. The body triplicates its initial $\ttt$,
and uses one copy as variant for an inner loop
that concatenates the other two copies.
\ee

\section{Programs for structure expansions}

In this section we describe programs that expand
arbitrary (finite) structures in important ways.

\subsection{Enumerators}\label{subsec:enumerator}
Given a $W$-structure \calS\ we say that a pair $(a,e)$, with
$a \in |\calS|$ and
$e: \; |\calS| \pa |\calS|$, is an {\em enumerator for \calS}
if for some $n$ the sequence
$$a,  e(a) ,  \ldots , e^{[n]}(a)$$
consists of all accessible nodes of \calS, without repetitions,
and $e^{[n+1]}(a)$ is undefined.
An enumerator is {\em monotone} if the value of a term never precedes
the value of its sub-terms.  This is guaranteed if the value of
a term of height $h$ never precedes the value of terms of
height $< h$.

\begin{thm}\label{thm:enumerator}
For each vocabulary $W$ there is a program that
for $W$-structures \calS\ as input yields an expansion of
$\calS$ with a monotone enumerator $E$.

\end{thm}
\prf
The program maintains, in addition to the identifiers in $W$,
four auxiliary identifiers, as follows.
\bi
\li A token \tta, intended to set the head $a$ of the enumerator.
\li A pointer \tte, intended to denote a (repeatedly growing) 
initial segment of the intended enumerator $e$;
\li A set identifier \ttE,
intended to denote the set of nodes enumerates by $e$ so far.
\li A pointer \ttd\ intended to list, starting from a token \ttb, some accessible nodes not
yet listed in \tte; these are to be appended to \tte\ at the end of each loop-cycle.
\li A token \ttf, intended to serve as a flag
to indicate that the last completed cycle
has added some elements to \ttE.
\ei

A preliminary program-segment sets \tta\ and \ttf\ to be the 
node denoted by one of the $V$-tokens
(there must be one, or else there would be no accessible nodes), 
and defines \tte\ to list any additional nodes denoted by tokens.
(The value of \ttf\ is immaterial, only \ttf\ being defined matters.)
Note that \tte, \ttd\ and \ttE\ are initially empty by default.

The main loop starts by re-initializing \ttd\ to empty, using
string duplication described above, resetting \ttf\
to undefined (i.e.\ false), and duplicating \tte\
as needed for the following construction.
Each pass then adds to \ttd\ all nodes that are obtained
from the current values in \tte\ by applications of $W$'s functions,
and that are not already in \ttE. That is, for each unary function-id \ttg\
of $W$ a secondary loop travels through \tte, using an auxiliary 
token $\ttt_1$.
When \ttg\ applied
to an entry is not in \ttE, the value of that output is appended to both
 \tta\ and \ttE.
The guard of that loop is $!\tte\ttt_1$, and the variant is \tte.

For function identifiers \ttg\ of arity $> 1$ the process is similar, except
that nested loops are required, with additional duplications of \tte\
ahead of each loop.
Whenever a new node is appended to \tta, the token \ttf\ is
set to be defined (say as the current vale of \tta).

When every non-nullary function-identifier of $W$ is treated, 
the list \tta\ is appended to \tte, leaving \tta\ empty.
\qed

\medskip

In \S \ref{subsec:examples} we gave a program for duplicating
a string. Using an enumerator, a program using the
same method would duplicate, for the accessible nodes,
each structure function.
Namely, to duplicate a $k$-ary function denoted by \ttf\ to one denoted
by $\ttf'$,
the program's traverses $k$ copies of the enumerator with $k$ tokens
$\ttc_1 \ldots \ttc_k$,
and whenever $\ttf\ttc_1 \cdots \ttc_k$ is defined,
the program defines $\ttf'\ttc_1 \cdots \ttc_k \sdownarrow \ttf\ttc_1 \cdots \ttc_k$.

Observe that an enumerator for a structure usually ceases to be one
with the execution of a structure revision; for example, a function
contraction may turn an accessible node into an inaccessible one.
This can be repaired by accompanying each revision by
an auxiliary program tailored to it, or simply by redefining an enumerator
whenever one is needed.

\subsection{Quasi-inverses}
We shall need to refer below to decomposition of inductive data,
i.e.\ inverses of constructors.  While in general structure functions
need not be injective, we can still have programs for quasi-inverses,
which we define as follows.\fn{A common equivalent definition is
that $f \circ g \circ f = f$.}

For a relation $R \subseteq A \times B$ and $a \in A$, define
$R'a \df \{b \in B \mid aRb\}$.\fn{We use infix notation for binary relations.}
We call a partial-function $f: \, A \pa B$
a {\em choice-function for $R$} if $f \subseteq R$
and $f(a)$ is defined whenever $R'a \neq \emptyset$.
A partial-function $g:\; A \sra B$ is a {\em quasi-inverse} of $f$
if it is a choice function for the relation $f^{-1}$.
When $f$ is $r$-ary, i.e.\ $A = \times_{i=1}^r A_i$,
$g$ can be construed as an $r$-tuple of functions $\lng g_1 \ldots g_r\rng$.
We write $f^{-i}$ for $g_i$.
If $f$ is injective then its unique quasi-inverse is its inverse $f^{-1}$.

\begin{thm}\label{thm:quasi-inverse}
For each vocabulary $W$ there is a program that
for each $W$-structure \calS\ as input yields an expansion of
$\calS$ with quasi-inverses for each non-nullary $W$-function.
\end{thm}
\prf
The proof of Theorem \ref{thm:enumerator} can be  easily modified
to generate quasi-inverses for each structure function, either
in tandem with the construction of an enumerator, or independently.
Namely, whenever the program in the proof of Theorem \ref{thm:enumerator}
adds a node $x= \ttg(x_1 .. x_k)$
to \tta\ and \ttE\ (where $k = \ar(\ttg)$), our enhanced program
defines $\ttg^{-i}(x)=x_i$ ($i=1..k$).
\qed

Note that, contrary to enumerators, quasi-inverses are easy to maintain
through structure revisions. An extension of a function \ttf\ 
can be augmented with appropriate extensions of \ttf's quasi-inverses, 
and a contraction of \ttf\ with appropriate contractions of those 
quasi-inverses. 

\section{A generic delineation of primitive recursion}

\subsection{Recurrence over inductive data}
Recall that the schema of {\em recurrence over \dN}
consists of the two equations
\begin{equation}\label{eq:Nrec}
\begin{array}{rcl}
f(0,\vec{x}) &=& g_0(\vec{x})\\
f(\tts n,\vec{x}) &=& g_s(n,\vec{x},f(n,\vec{x}))
\end{array}
\end{equation}
 
More generally, given a free algebra $\dA = \dA(C)$ generated from 
a finite set
$C$ of constructors,
{\em recurrence over \dA} has one equation per constructor:
\begin{equation}\label{eq:Crec}\begin{array}{rcl}
f(\ttc(z_1, \ldots , z_k),\vec{x}) 
	&=& g_{\scriptttc}(\vec{x},\vec{z},y_1 \ldots y_k)\\
	\text{where} \quad
		y_j &=& f(z_j,\vec{x}) \quad (j=1..k, \; k = \ar(\ttc))
\end{array}\end{equation}
The set $\bxbf{PR}(\dA)$  of {\em primitive recursive functions} over \dA\
is generated from the constructors of \dA\ (for example zero and successor for \dN), 
by recurrence over \dA\ and explicit definitions.\fn{The phrase
``primitive recursive'' was coined by Rosza Peter \cite{Peter51}, triggered by 
the discoveries by Ackermann and Sudan of computable (``recursive'') functions
that are not in $\bxbf{PR}(\dN)$.  Given the present-day 
use of ``recursion'' for recursive procedures, ``recurrence'' seems all the
more appropriate.}
Using standard codings, it is easy to see that
any non-trivial (i.e.\ infinite) algebra
can be embedded in any other.
Consequently, the classes $\bxbf{PR}(\dA)$
are essentially the same for all
non-trivial \dA, and we refer to them jointly as \bxbf{PR}.\fn{Note
that we are {\em not} dealing in generalizations of
recurrence to well-orderings (``Noetherian induction'').}
A natural question is whether there is a generic approach, unrelated
to free algebras, that delineates the class \bxbf{PR}.

The recurrence schema (for \dN) was seemingly initiated by the 
interest of Dedekind in formalizing arithmetic, and articulated by Skolem 
\cite{Skolem23}.
It was studied extensively (e.g.\ \cite{Peter51}),
and generalized to all admissible structures \cite{Barwise-admissible}.
Our aim here is to characterize the underlying notion of primitive recursion 
generically, via uninterpreted programs.  
We delineate a natural variant
of \bxbf{ST}, \bxbf{STV} which is sound and complete for
\bxbf{PR}. 
That is, on the one hand every \bxbf{STV} program terminates in time
primitive-recursive in the size of the input structure.
On the other hand, \bxbf{STV} captures \bxbf{PR} in two ways:
any instance of recurrence over a free algebra
can be implemented directly by an \bxbf{STV} program; and
every \bxbf{ST} program that runs in \bxbf{PR} resources in the
size of the input structure can be transformed into an
extensionally equivalent \bxbf{STV} program.

Recurrence is guaranteed to terminate because it consumes
its recurrence argument. The very same consumption phenomenon is used,
in a broad and generic sense, in
the Dijkstra-Hoare program verification style, 
in the notion of a {\em variant} \cite{Gries81,Dijkstra76,Winskel93}.  
Our core idea is to use a generic notion of program variants
in lieu of recurrence arguments taken from free algebras.


\subsection{Resource measures}\label{subsec:resource}


We first identify appropriate notions of size
measures for structures. We focus on accessible structures,
since non-accessible nodes
remain non-accessible through revisions and are
inert through the execution of any program.
Consequently they do not affect the time or space consumption of computations.

We take the {\em size} $\#\calS$ of an accessible $V$-structure \calS\ to be
the count of tuples of nodes that occur in the structure's
relations and (graphs of) functions.
Note that this is in tune with our use of variants,
which are consumed not by the elimination of nodes, but by the
contraction of functions and relations.
Moreover, we believe that the size of functions and relations
is an appropriate measure in general, since they convey more accurately
than the number of nodes the information contents of a structure.

Note that for word-structures, i.e. $\calT(w)$ for $w \in \grS^*$
(\grS\ an alphabet) the total size of the structure's functions
is precisely the length or $w$, so in this important case our measure
is identical to the count of nodes.

Suppose $V$ is a vocabulary with all
identifiers of arity $\leq r$.
If \calS\ is a $V$-structure of size $k$,
then the number of accessible nodes is $O(k^r)$.  
Conversely, if the number of accessible nodes is $a$,
then the size is $O(a^{r+1})$.
It follows that the distinction between our measure and 
node-count does not matter for super-polynomial complexity.

We say that a program $P$ runs {\em within time} $t: \dN \sra \dN$ if
for all structures \calS, the number of configurations in a 
complete trace of $P$ on input \calS\ is $\leq t(\# \calS)$;
it runs {\em within space} $s: \dN \sra \dN$ if
for all \calS, all configurations in an execution
trace of $P$ on input \calS\ are of size $\leq s(\# \calS)$.

We say that $P$ {\em runs in PR} if
it runs within time $t$, for some PR function $t$, or --- equivalently ---
within space $s$, for some PR function $s$.

\subsection{PR-soundness of \bxbf{STV}-programs}

We assign to each \bxbf{STV}-program $P$ a primitive-recursive
function $b_P: \, \dN \sra\dN$ as follows.
The aim is to satisfy Theorem \ref{thm:PR-bounds} below.
\bi
\li If $P$ is an extension or an inception revision,
then $b_P(n) = 1$; if $P$ is any other revision then $b_P(n) = 0$.
\li If $P$ is $S;Q$ then $b_P(n) = b_Q(b_S(n))$
\li If $P$ is $\bxbf{if}[G]\{S\}\{Q\}$ then
$b_P(n) = \max[b_S(n),b_Q(n)]$.
\li If $P$ is $\bxbf{do}[G][T]\{Q\}$ then
$b_P(n) = b_Q^{[n]}(n)$.
\ei

\begin{thm}\label{thm:PR-bounds}
If $P$ is an \bxbf{STV}-program computing a mapping $\grF_P$ between
structures, and \calS\ is a structure, then
$$
\# \grF_P(\calS) \leq b_P(\# \calS)
$$
\end{thm} 
\prf
Structural induction on $P$.
\bi
\li
If $P$ is a revision, then the claim is immediate by 
the definition of $b_P$.
\li
If $P$ is $S;Q$ then 
\beqnaa
\# \grF_P(\calS) &=& \# \grF_Q(\grF_S(\calS)) \\
	& \leq & b_Q(\# \grF_S(\calS)) & \text{(IH for $Q$)}\\
	& \leq & b_Q(b_S(\# \calS)) 
		& \text{(IH for $S$, $b_Q$ is non-decreasing)}\\
	& = & b_P(\# \calS)
\eeqnaa
\li 
The case for $P$ of the form $\bxbf{if}[G]\{S\}\{Q\}$ is immediate.
\li
If $P$ is $\bxbf{do}[G][T]\{Q\}$ then
$\grF_P(\calS)$ is $\grF_Q^{[m]}(\calS)$ for some $m$.
By the definition of variants, and the semantics of looping,
$m$ is bounded by the size of $T$, which is bounded by the
size of $\calS$. So
\beqnaa
\# \grF_P(\calS) &=& \# \grF_Q^{[m]}(\calS) 
	& \text{for some $m \leq \# \calS$} \\
		&\leq & b_Q^{[m]}(\# \calS)
	& \text{IH, $b_Q$ is non-decreasing} \\
		&\leq & b_Q^{[n]}(\# \calS)
	& \text{where $n = \# \calS$} \\
	&&& \zero\quad \text{since $b_Q$ is terminating} \\
		& = & b_P(\# \calS)
\eeqnaa
\ei
\qed

From Theorem \ref{thm:PR-bounds} we obtain the soundness of
\bxbf{STV}-programs for PR:

\begin{thm}
Every \bxbf{STV}-program runs in PR space, and therefore in PR time.
\end{thm}

\subsection{Completeness of \bxbf{STV}-programs for PR}

We finally turn to the completeness of \bxbf{STV} for \bxbf{PR}.
The easiest approach would be to prove that \bxbf{STV} is complete
for $\bxbf{PR}(\dN)$, and then invoke the coding of primitive recurrence
over any free algebra in $\bxbf{PR}(\dN)$.  This, however,
would fail to establish a direct representation of generic recurrence 
by \bxbf{STV}-programs, which is
one of the {\em raisons d'\^{e}tre} of \bxbf{STV}.
We therefore follow a more general approach.

\begin{lem}\label{lem:complete-for-rec}
For each free algebra $\dA(C)$, each instance 
of recurrence over \dA\ as in (\ref{eq:Crec}) above 
(with $\vec{x} = x_1, \ldots x_m$), the following holds.
Given \bxbf{STV}-programs for the functions
$g_{\scriptttc}$, there is an \bxbf{STV}-program $P$ that maps 
the structure $\calT(w) \oplus \calT(x_1) \oplus \cdots \oplus \calT(x_m)$
to $\calT(f(w,x_1, \ldots ,x_m))$.
\end{lem}
\prf
The program $P$ gradually constructs a pointer 
\ttr\ that maps each node \grn\ of
$\calT(w)$ to the root of the structure $\calT(f(u,\vec{x}))$,
where $u$ is the sub-term of $w$ denoting \grn\ (it is uniquely defined
since $\dA$ is a free algebra).

$P$ starts by constructing a monotone enumerator for the structure $\calT(w)$,
as well as inverses for all constructors, by Theorems \ref{thm:enumerator}
and \ref{thm:quasi-inverse}. (Since $w$ is a term of a free algebra,
a quasi-inverse of a constructor is an inverse).
The main loop of $P$ then scans that enumerator, using a token;
reaching the end of the enumerator is the guard, and the
enumerator itself is the variant.

For each node \grn\ encountered on the enumerator, 
$P$ first identifies the constructor \ttc\ defining \grn, which is 
unique since $w \in \dA(C)$.  
This identification is possible by testing for equality
with the tokens, and --- that failing --- testing, 
for non-nullary constructor
\ttf, the definability of the first inverse $\ttf^{-1}$. 
Since the enumerator is monotone, \ttr\ is already defined for the values
$z_1\! =\! \ttf^{-1}(\grn),\; \ldots,\; z_k \!=\! \ttf^{-k}(\grn)$ 
($k = \ar(\ttc)$).
$P$ can thus invoke the program $P_{\fnttc}$ for the function $g_{\fnttc}$,
adapted to the disjoint union of
\be
\li The structures $\calT(x_i))$;
\li The structures spanned by the $z_j$'s,
i.e.\ for each $j$ the substructure of the input consisting of
the sub-terms of $z_j$;
\li The structures $\ttr(z_j)$ already obtained.
\ee
$\ttr(\grn)$ is then set to be the root of the result.

The program's final output is then $\ttr\,\bullet$; that is
the structure yielded for the program's given recurrence argument.
\qed

\begin{thm}\label{thm:complete-for-PR}
For each free algebra $\dA$,  
the collection of \bxbf{STV}-programs is complete for $\bxbf{PR}(\dA)$.
\end{thm}
\prf
The proof proceeds by induction on the PR definition of $f$.
The cases where $f$ is a constructor are trivial. 
For explicit definitions, and more particularly composition,
we need to address the need of duplicating substructures,
for which we have programs, as explained in \S \ref{subsec:enumerator}.

Finally, the case of recurrence is treated in Lemma \ref{lem:complete-for-rec}.
\qed

Theorem \ref{thm:complete-for-PR} establishes a simple and direct
mapping of PR function definitions, over any free algebra, to 
\bxbf{STV} programs.
Another angle on the completeness of \bxbf{STV} for \bxbf{PR} refers
directly to \bxbf{ST}-programs (i.e.\ to programs without variants):

\begin{cor}\label{cor:compl-for-pr-resources}
For every \bxbf{ST}-program $P$ running in PR resources, and defining
a structure transformation \grF, there is an \bxbf{STV}-program $Q$ 
that computes \grF.
\end{cor}
\prf
Recall from \S\ref{subsec:resource} that the size of a structure,
measured in size of functions and relations, is polynomial in the
number of nodes. It follows that $P$ runs in time PR in
the input's number of nodes.

Suppose now that $P$'s input is a $V$-structure,
and that $P$ operates within time $f(n)$, where $f$ is a PR function
over \dN. 

Let $Q$ be the composition of the following \bxbf{STV}-programs:
\be
\li A program
that expands each $V$-structure \calS\ with an enumerator $(a,e)$,
as in Theorem \ref{thm:enumerator}.  The constructed enumerator \tte\
is a list without repetition of the nodes of \calS.
I.e., \tte\ is essentially $\calT(n)$, 
where $n$ is the number of nodes in \calS.
\li A program that 
takes as input the structure $\calT(n)$ constructed in (1),
and outputs $\calT(f(n))$ with, say, \ttt\ as
the output's successor function.  Such a program exists by Theorem
\ref{thm:complete-for-PR} applied to the free algebra \dN.
\li The given \bxbf{ST}-program $P$, 
with each loop assigned as variant a copy of \ttt, 
and each loop-body preceded by a function-contraction of \ttt.
\ee
Then $Q$ computes the same structure-transformation as $P$.
\qed
 
%
%

\mywhite{\cite{Heijenoort67}}
\bibliographystyle{plain}
\bibliography{x}
\end{document}

%% file: all.tex
\usepackage{stmaryrd}

\newcommand{\ar}{{\mathfrak r}} 
\newcommand{\hgt}{{\mathfrak h}} 

\newcommand{\pfont}{\sffamily\bfseries}
\newcommand{\save}[1]{}

\renewcommand{\leq}{\leqslant}
\renewcommand{\geq}{\geqslant}

\newcommand{\li}{\item\zero}

\newcommand{\ul}[1]{\underline{#1}}

\newcommand{\ara}[1]{{\stackrel{#1}{\longrightarrow}}} 

\newcommand{\dara}[2]%
      {\overset{#1}{\underset{\raisebox{4ex}{\;\;\fnit{#2}}}{\longrightarrow}}} 

\newcommand{\myblue}[1]{\NavyBlue{#1}}

\newcommand{\mywhite}[1]{\White{#1}}

\newcommand{\outt}[1]{}
\newcommand{\delete}[1]{}
\newcommand{\blank}[1]{}

\newcommand{\amp}{\;\;\;\hbox{\scriptsize$\&$}\;\;\;}
\newcommand{\samp}{\;\hbox{\scriptsize$\&$}\;}
\newcommand{\notedomission}[1]{\medskip\noindent{\bf TEXT OMITTED}\\[2mm]}

\newenvironment{xtr}{\huge\tt\hspace{-5mm}}{\\[2mm] \zero\hrulefill} 
\newcommand{\bxtr}{\begin{xtr}\begin{envviolet}\noindent}
\newcommand{\extr}{\end{envviolet}\end{xtr}}

\newcommand{\bmini}{\begin{minipage}[t]{0.8\textwidth}}
\newcommand{\emini}{\end{minipage}}

\newenvironment{envviolet}{\textViolet}{\textBlack}

\newcommand{\bxit}[1]{\hbox{\it #1}}

\newcommand{\bxrm}[1]{\hbox{\rm #1}}

\newcommand{\bxbf}[1]{\hbox{\bf #1}}

\newcommand{\bxsc}[1]{\hbox{\sc #1}}
\newcommand{\bxtt}[1]{\hbox{$\tt #1$}}

\newcommand{\bxsl}[1]{\hbox{\sl #1}}
\newcommand{\fnbf}[1]{\hbox{\footnotesize\bf #1}}
\newcommand{\fnit}[1]{\hbox{\footnotesize\it #1}}

\newcommand{\fnrm}[1]{\hbox{\footnotesize #1}}
\newcommand{\fntt}[1]{\hbox{\footnotesize\tt #1}}

\newcommand{\ttscript}[1]{\hbox{\scriptsize\bf #1}}

\newcommand{\scriptbf}[1]{\hbox{\scriptsize\bf #1}}
\newcommand{\scripttt}[1]{\hbox{\scriptsize\bf #1}}

\newcommand{\scriptttc}{\hbox{\scripttt{c}}}

\newcommand{\fnbfq}{\fnbf{q}}

\newcommand{\fnbfR}{\fnbf{R}}

\newcommand{\fnttc}{\fntt{c}}

\newcommand{\bfc}{\hbox{\bf c}}

\newcommand{\bff}{\hbox{\bf f}}

\newcommand{\bfo}{\hbox{\bf o}}

\newcommand{\bfq}{\hbox{\bf q}}

\newcommand{\bfs}{\hbox{\bf s}}
\newcommand{\bft}{\hbox{\bf t}}

\newcommand{\sla}{\hbox{$\sl a$}}
\newcommand{\slb}{\hbox{$\sl b$}}
\newcommand{\slc}{\hbox{$\sl c$}}
\newcommand{\sld}{\hbox{$\sl d$}}

\newcommand{\tta}{\hbox{$\tt a$}}
\newcommand{\ttb}{\hbox{$\tt b$}}
\newcommand{\ttc}{\hbox{$\tt c$}}
\newcommand{\ttd}{\hbox{$\tt d$}}
\newcommand{\tte}{\hbox{$\tt e$}}
\newcommand{\ttf}{\hbox{$\tt f$}}
\newcommand{\ttg}{\hbox{$\tt g$}}
\newcommand{\tth}{\hbox{$\tt h$}}
\newcommand{\tti}{\hbox{$\tt i$}}

\newcommand{\ttk}{\hbox{$\tt k$}}
\newcommand{\ttl}{\hbox{$\tt l$}}
\newcommand{\ttm}{\hbox{$\tt m$}}
\newcommand{\ttn}{\hbox{$\tt n$}}

\newcommand{\ttp}{\hbox{$\tt p$}}
\newcommand{\ttq}{\hbox{$\tt q$}}
\newcommand{\ttr}{\hbox{$\tt r$}}
\newcommand{\tts}{\hbox{$\tt s$}}
\newcommand{\ttt}{\hbox{$\tt t$}}
\newcommand{\ttu}{\hbox{$\tt u$}}
\newcommand{\ttv}{\hbox{$\tt v$}}
\newcommand{\ttw}{\hbox{$\tt w$}}

\newcommand{\tty}{\hbox{$\tt y$}}
\newcommand{\ttz}{\hbox{$\tt z$}}

\newcommand{\ttzero}{\hbox{$\tt 0$}}

\newcommand{\ttone}{\hbox{$\tt 1$}}

\newcommand{\bfI}{\hbox{\bf I}}

\newcommand{\bfN}{\hbox{\bf N}}

\newcommand{\bfQ}{\hbox{\bf Q}}
\newcommand{\bfR}{\hbox{\bf R}}

\newcommand{\gothC}{\mbox{$\mathfrak C$}}

\newcommand{\bfzero}{{\bf 0}}

\newcommand{\calQ}{\hbox{$\cal Q$}}
\newcommand{\calR}{\hbox{$\cal R$}}
\newcommand{\calS}{\hbox{$\cal S$}}
\newcommand{\calT}{\hbox{$\cal T$}}

\newcommand{\tinycalQ}{\hbox{{\tiny $\cal Q$}}}
\newcommand{\tinycalR}{\hbox{{\tiny $\cal R$}}}
\newcommand{\tinycalS}{\hbox{{\tiny $\cal S$}}}
\newcommand{\tinycalT}{\hbox{{\tiny $\cal T$}}}

\newcommand{\ttA}{\hbox{$\tt A$}}

\newcommand{\ttC}{\hbox{$\tt C$}}
\newcommand{\ttD}{\hbox{$\tt D$}}
\newcommand{\ttE}{\hbox{$\tt E$}}

\newcommand{\ttI}{\hbox{$\tt I$}}

\newcommand{\ttL}{\hbox{$\tt L$}}
\newcommand{\ttM}{\hbox{$\tt M$}}

\newcommand{\ttR}{\hbox{$\tt R$}}

\newcommand{\dA}{\hbox{$\Bbb A$}}

\newcommand{\dN}{\hbox{$\Bbb N$}}

\newcommand{\gra}{\hbox{$\alpha$}}
\newcommand{\grb}{\hbox{$\beta$}}
\newcommand{\grg}{\hbox{$\gamma$}}
\newcommand{\grd}{\hbox{$\delta$}}
\newcommand{\gre}{\hbox{$\varepsilon$}}

\newcommand{\grn}{\hbox{$\nu$}}

\newcommand{\grr}{\hbox{$\rho$}}
\newcommand{\grs}{\hbox{$\sigma$}}
\newcommand{\grt}{\hbox{$\tau$}}

\newcommand{\grf}{\hbox{$\varphi$}}
\newcommand{\grch}{\hbox{$\chi$}}

\newcommand{\grG}{\hbox{$\Gamma$}}

\newcommand{\grS}{\hbox{$\Sigma$}}

\newcommand{\grF}{\hbox{$\Phi$}}
\newcommand{\grQ}{\hbox{$\Psi$}}

\newcommand{\bgrs}{\hbox{\boldmath$\sigma$}}

\newcommand{\hd}{\bxsf{hd}}

\newcommand{\ra}{\rightarrow}
\newcommand{\pa}{\rightharpoonup}
\newcommand{\sra}{\!\rightarrow\!} 

\newcommand{\rA}{\Rightarrow}  

\newcommand{\stimes}{\!\times\!}
\newcommand{\splus}{\!+\!}
\newcommand{\sminus}{\!-\!}

\newcommand{\suparrow}{\neghalfmm\uparrow\neghalfmm}
\newcommand{\sUparrow}{\neghalfmm\Uparrow\neghalfmm}

\newcommand{\sdownarrow}{\neghalfmm\downarrow\neghalfmm}
\newcommand{\sDownarrow}{\neghalfmm\Downarrow\neghalfmm}
\newcommand{\ssDownarrow}{\negonemm\Downarrow\negonemm}

\newcommand{\sequal}{\!=\!}
\newcommand{\seq}{\!=\!}
\newcommand{\sneq}{\!\neq\!}

\newcommand{\qed}{\hfill {\boldmath$\Box$}\\}
\newcommand{\lng}{\langle}
\newcommand{\rng}{\rangle}
\newcommand{\df}{=_{\rm df}}

\newcommand{\rsem}{\,]\hspace{-0.6mm}]} 
\newcommand{\lsem}{[\hspace{-0.6mm}[\,} 

\newcommand{\ignore}[1]{}

\newcommand{\mx}{\makebox}

\newcommand{\zero}{\rule{0mm}{3mm}}

\newcommand{\negonemm}{\mx[-1mm]{}}
\newcommand{\neghalfmm}{\mx[-0.5mm]{}}

\newcommand{\threemm}{\mx[3mm]{}}

\newcommand{\onecm}{\mx[1cm]{}}

\newcommand{\bc}{\begin{center}}
\newcommand{\ec}{\end{center}}
\newcommand{\beq}{\begin{equation}}
\newcommand{\eeq}{\end{equation}}

\newcommand{\be}{\begin{enumerate}}
\newcommand{\ee}{\end{enumerate}}
\newcommand{\bi}{\begin{itemize}}
\newcommand{\ei}{\end{itemize}}
\newcommand{\bd}{\begin{description}}
\newcommand{\ed}{\end{description}}
\newcommand{\beqn}{\begin{equation}}
\newcommand{\eeqn}{\end{equation}}

\newcommand{\beqna}{\begin{eqnarray}}
\newcommand{\eeqna}{\end{eqnarray}}
\newcommand{\beqnas}{\begin{eqnarray*}}
\newcommand{\eeqnas}{\end{eqnarray*}}
\newcommand{\beqnaa}{$$\begin{array}{rcll}}  
\newcommand{\eeqnaa}{\end{array}$$}  
\newcommand{\beqnal}{$$\begin{array}{l}}  
\newcommand{\eeqnal}{\end{array}$$}  
\newcommand{\beqnana}{$$\begin{array}{lrcll}}  
\newcommand{\eeqnana}{\end{array}$$}  
\newcommand{\btbl}[1]{\begin{center}\begin{tabular}{#1}}
\newcommand{\etbl}{\end{tabular}\end{center}}
\newcommand{\beqnc}{$$\begin{array}{rclcl}}
\newcommand{\eeqnc}{\end{array}$$}
\newcommand{\fn}{\footnote}

\newcommand{\prf}{{\sc Proof. }}

\newtheorem{dclprop}{{\sc Proposition}} 
\newtheorem{dclprops}{{\sc Proposition}}[subsection] 
\newtheorem{dclbigthm}[dclprop]{THEOREM}

\makeatletter
\def\thmlabel#1{\@bsphack\if@filesw {\let\thepage\relax
\xdef\@gtempa{\write\@auxout{\string
\newlabel{#1}{{\@Roman{\@currentlabel}}{\thepage}}}}}\@gtempa
\if@nobreak \ifvmode\nobreak\fi\fi\fi\@esphack}
\makeatother

\newtheorem{dclthm}[dclprop]{{\sc Theorem}}   
\newtheorem{dclthms}[dclprops]{{\sc Theorem}}   

\newtheorem{dcllem}[dclprop]{{\sc Lemma}}
\newtheorem{dcllems}[dclprops]{{\sc Lemma}} 
\newtheorem{dclsublem}[dclprop]{{\sc Sublemma}}
\newtheorem{dclcor}[dclprop]{{\sc Corollary}}
\newtheorem{dclcors}[dclprops]{{\sc Corollary}} 
\newtheorem{dcldfn}[dclprop]{{\sc Definition}}
\newtheorem{dcldfns}[dclprops]{{\sc Definition}}
\newtheorem{dclasss}[dclprops]{{\bf Assumption}}
\newtheorem{dclass}[dclprop]{{\bf Assumption}}

\newenvironment{prop}{\medskip\begin{dclprop}\sl}{\end{dclprop}}
\newenvironment{thm}{\medskip\begin{dclthm}\sl}{\end{dclthm}}

\newenvironment{cor}{\medskip\begin{dclcor}\sl}{\end{dclcor}}

\newenvironment{dfn}{\medskip\begin{dcldfn}\sl}{\end{dcldfn}}

\newenvironment{lem}{\medskip\begin{dcllem}\sl}{\end{dcllem}}

\newcommand{\bsl}{\begin{verse}\sl}
\newcommand{\esl}{\end{verse}}

\newtheorem{ex}[dclprop]{Example}
\newtheorem{exxs}[dclprop]{Exercises}

\newenvironment{exercises-with-preamble}{\begin{exxs}\rm}{\end{exxs}}

\newcommand{\bthm}{\begin{thm}}
\newcommand{\ethm}{\end{thm}}
\newcommand{\bprop}{\begin{prop}}
\newcommand{\eprop}{\end{prop}}
\newcommand{\blem}{\begin{lem}}
\newcommand{\elem}{\end{lem}}
\newcommand{\bcor}{\begin{cor}}
\newcommand{\ecor}{\end{cor}}
\newcommand{\bdfn}{\begin{dfn}}
\newcommand{\edfn}{\end{dfn}}

\newcommand{\bz}{\begin{quote}\small}
\newcommand{\ez}{\end{quote}}

\newcommand{\einference}[2]  
  {\shortstack
      {$ #1 $\\ \mbox{}\\ $ #2 $}}

\newlength{\txtlth}
\newlength{\txtht}

